\documentstyle[aps,twocolumn,epsf]{revtex}
\draft
\newcommand{\ave}[1]{\left\langle #1\right\rangle}
\newcommand{\myref}[1]{(\ref{#1})}
\newcommand{\figsize}{3.0 in}
\newcommand{\gammacorr}{\tau_c}            %C(t)\sim t^{-\gammacorr}
\newcommand{\taucorr}{\sigma_c}          %moment exponent for C(t,L)
\newcommand{\fcorr}{f_{c}}               %spectrum for C(t,L)
\newcommand{\tausize}{\sigma_s}          %moment exponent for P(s,L)
\newcommand{\tauwaves}[1]{\sigma_{s,#1}} %moment exponent for P(s,n,L)
\newcommand{\fwaves}{f_{w}}              %spectrum for P_w(s,L)

\begin{document}
\title{From waves to avalanches: two different mechanisms of sandpile 
dynamics}
\author{Mario De Menech$^1$ and Attilio L. Stella$^{1,2}$}
\address{$^1$ INFM-Dipartimento di Fisica,\\
Universit\`a di Padova, I-35131 Padova, Italy\\
$^2$ The Abdus Salam ICTP, P. O. Box 563, I-34100, Trieste, Italy,\\
Sezione INFN, Universit\`a di Padova, I-35131 Padova, Italy }
\date{\today} 
\maketitle

\begin{abstract}
Time series resulting from wave decomposition
show the existence of different correlation
patterns for avalanche dynamics. For the $d=2$ Bak-Tang-Wiesenfeld
model, long range correlations determine a modification of
the wave size distribution under coarse graining in time, and
multifractal scaling for avalanches.
In the Manna model, the 
distribution of avalanches coincides with that of waves, 
which are uncorrelated and obey finite size scaling, a
result expected also for the $d=3$ Bak et al. model.

\bgroup
\egroup
\end{abstract}

%\vspace{1cm} 
%\narrowtext 

Full information  on a self--organized critical (SOC)
process~\cite{BTW} is contained in the time series, if the time step
is the most microscopic conceivable. The self-similarity of such a
process, due to its intermittent, avalanche character,
 should be revealed by the scaling 
of the time autocorrelation or its power spectrum. 
In spite of this, time series analyses have been seldom
performed on sandpiles or similar systems, and mostly concerned
the response to finite, random external disturbances,
i.e. the problem of $1/f$ noise~\cite{1/f}.
 Most efforts
in the characterization of SOC scaling 
concentrated on probability distribution functions (PDF's) of global
properties of the avalanche events, which occupy large intervals of the
microscopic evolution time~\cite{PDFs}.
 The numerical analysis of such PDF's is often 
difficult, and universality issues can not be 
easily solved.
The situation is even more problematic if, like for
the two-dimensional (2D) Bak-Tang-Wiesenfeld sandpile (BTW)~\cite{BTW},
the usually assumed finite size scaling (FSS) form reveals inadequate
for the PDF's, and needs to be replaced by a multifractal one~\cite{Tebaldi}. 
These findings rise 
the additional issue of why the 2D BTW displays such multifractal scaling, 
while apparently very similar sandpiles, like the Manna model
(M)~\cite{Manna-model},
do not~\cite{Tebaldi,Chessa,Lubeck}. 

In recent theoretical approaches to the BTW and other 
Abelian sandpiles~\cite{Dhar-ADP}, 
a major role has been played by the waves of 
toppling into which avalanches can be decomposed~\cite{Ivashkevich}. 
For the BTW, the PDF's of waves, as sampled from a large collection of
successive avalanches, obey FSS with exactly known 
exponents~\cite{Ktitarev-Lubeck}. 
By analyzing the succession of waves as a 
stationary time series, one could hope to determine the statistical properties 
of avalanches. However, so far, no precise information on
the correlations of such series was obtained. 

In this Letter we generalize the wave description to the M 
in 2D. The study of the respective wave time series
reveals that BTW and M are prototypes of two very different scenarios
for avalanche dynamics. In the M case, successive wave sizes
are totally uncorrelated. As a consequence, avalanche and wave PDF's
have identical scaling properties, consistent with FSS.
For the BTW, to the contrary, wave sizes show long range correlations
and persistency in time. 
PDF's of ``block variables'' given by sums of $n$ 
successive wave sizes show that these correlations are responsible
for the fact that avalanche scaling differs from the wave one and
has multifractal features. 
For the 3D BTW, on the other hand, our results suggest validity of
a scenario identical to the M one, and lead to conjecture exact avalanche
exponents coinciding with those of the wave PDF.

The 2D BTW~\cite{BTW} is defined on a square 
$L\times L$ lattice. An integer
$z_i$, the number of grains, is assigned to  site $i$.
Starting from a stable
configuration ($z_i\leq z_c=4$,  $\forall i$), a grain is added 
at a randomly 
chosen site; after each addition, all sites exceeding the stability 
threshold, $z_k>z_c$,
undergo toppling, distributing one grain to each one of the nearest neighbors.
The topplings, which dissipate grains when occurring at the edges, continue
until all sites are stable, and a new grain is added.
The $s$ topplings  between two consecutive additions 
form an avalanche. After many
additions, the system organizes into a stationary critical state.

Manna~\cite{Manna-model} studied a two--state version, M, of the sandpile.
The sites can be either empty or occupied;
grains are added randomly, and when one of them drops onto an
occupied site, a ``hard core'' repulsion pushes two particles out to 
randomly chosen nearest neighbors. Compared to the BTW,
in which toppling is deterministic, this model has
an extra stochastic ingredient in the
microscopic evolution, and $z_c=1$.

One can define  a special ordering in
the topplings of the BTW by introducing the wave decomposition of
avalanches~\cite{Ivashkevich}. 
After the site of addition, $O$, topples,
any other unstable site is allowed to topple except possibly $O$. 
This is the  first wave of toppling. If $O$ is still unstable, it
is allowed to topple once again, leading to the
second wave. This continues until $O$ becomes
stable. Thus, an avalanche is broken into a sequence of waves.
During a wave,  sites
topple only once, and for an $m$-wave avalanche $s=\sum_{k=1}^m s_k$,
where $s_k$ are the topplings of the $k$-th wave.
Following the definitions for the BTW, 
we implement here a wave decomposition of avalanches 
also for the M. Unlike for the BTW, sites involved
in a wave may topple more than once.
Furthermore, the toppling order chosen implies
now a peculiar sequence
of stable configurations visited upon addition of grains, but
the realization probabilities of possible 
 configurations are independent of this order~\cite{Dhar-ADP}.

The BTW wave size PDF has FSS form
$P_w(s)\sim s^{-\tau_w}p_w(s/L^{D_w})$, with $\tau_w=1$,  
$D_w=2$ in 2D, and $p_w$ a suitable scaling 
function~\cite{Ivashkevich,Ktitarev-Lubeck}.
For the M waves we also found a PDF obeying FSS, with
$\tau_w=1.31 \pm 0.02$ and $D_w=2.75 \pm 0.01$. These
exponents
 are remarkably consistent with those estimated for
avalanches~\cite{Manna-model,Chessa}. 
Such a coincidence certainly does not apply to the BTW.
Attempts to derive exact
BTW avalanche exponents within FSS
were based on the observation that often waves within
an avalanche show a final, long contraction phase, and on a 
scaling assumption for the corresponding
$s_k-s_{k+1}$~\cite{Priezzhev}.
 A more adequate ansatz for the conditional PDF
of $s_{k+1}$ given $s_k$~\cite{Paczuski},
and Markovianity assumptions,
did not help in better characterizing avalanche scaling
along these lines~\cite{Ktitarev-waves}.
 In fact the 2D BTW obeys a multifractal
form of scaling, which is not catched by such simplified
schemes~\cite{Tebaldi}.

The wave time series $\{s_k\}$
provide coarse grained dynamical descriptions.
In the $L\to \infty$ limit, these descriptions
are infinitely rescaled with respect to those at microscopic time scale, but
still infinitely finer than the mere records of successive
avalanche sizes. This intermediate 
time scale reveals essential in order to understand the dynamics
indside avalanches, whose size sequence we found to be 
uncorrelated in the sense discussed below for waves. 
The microscopic scale gave no
significant results in  comparing  the two models, since,
at that level, similar strong correlations exists in both of them,
due to the parallel updating algorithm.

\begin{figure}[tbp]
  \centerline{
  \epsfxsize=\figsize
  \epsffile{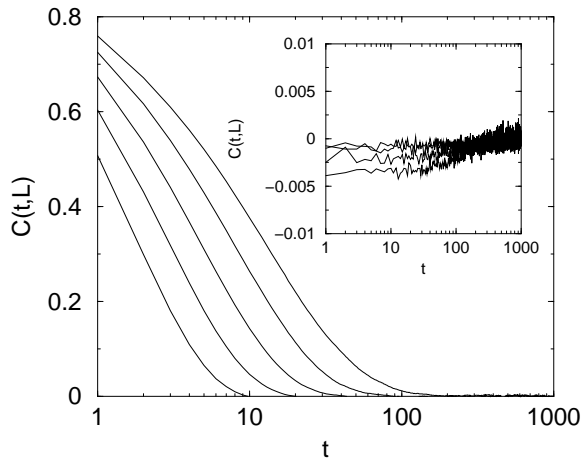}
  }
  \caption{For the  BTW model, as $L$ grows, $C$ decays more and more slowly,
  ($L=128,256,512,1024,2048$ from left to right). $C(t\geq 1,L)\simeq 0$ 
 for the  M (inset).}
  \label{fig:corrarea}
\end{figure}

\begin{figure}[tbp]
  \centerline{
  \epsfxsize=\figsize
  \epsffile{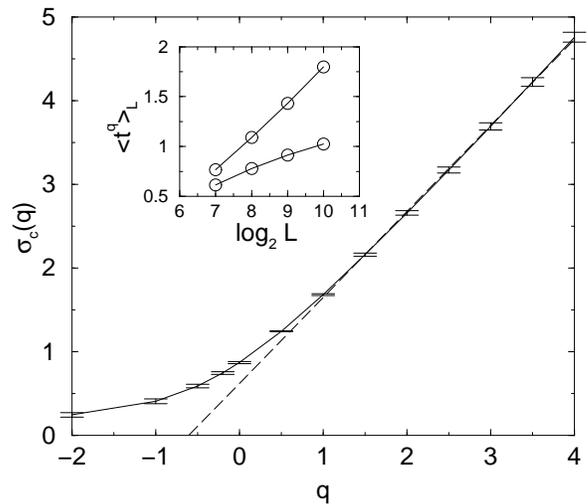}
  }
  \caption{The
  dashed curve interpolates $\taucorr(q)$ for $1\leq q\leq 4$; it has slope
  $D=1.02$ and intercept $0.62$. The inset shows  semi--logarithmic
  plots of $\ave{t^q}_L$ vs $\log_2 L$ for $q=-1,-0.5$.}
  \label{fig:tauqcorr}
\end{figure}

\begin{figure}[htb]
  \centerline{
  \epsfxsize=\figsize
  \epsffile{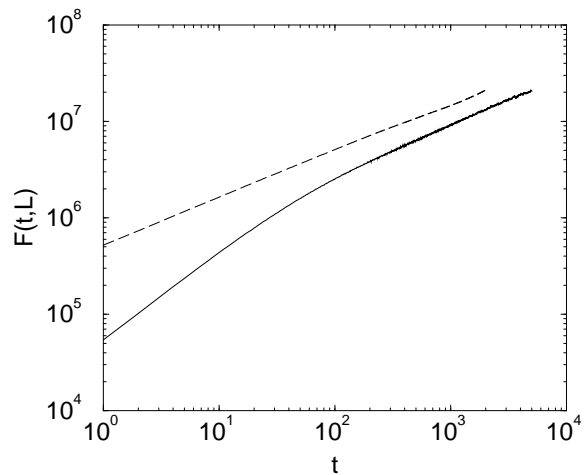}
  }
  \caption{Plots of $F(t,L=2048)$ for the BTW (continuos) and the M 
(dashed, slope $=0.5$).} 

  \label{fig:variance}
\end{figure}

We determined for BTW and M the autocorrelation 
\begin{equation}
C(t,L)=\frac{\ave{s_{k+t}s_{k}}_L-\mu^2}{\ave{s^2_k}_L-\mu^2},
\end{equation}
with $t=1,2,\ldots$, and $\mu=\ave{s_k}_L$,  the time averages being taken
over up to $10^7$
 waves for  $L=128,256,512,1024$ and $2048$. 
A first, striking result is that, as soon as $t>0$ ($C(0,L)=1$ by
normalization),
 waves are uncorrelated in the M. Indeed, as $L$ grows,
$C$ manifestly approaches $0$ as soon
as $t>0$ (Fig.~\ref{fig:corrarea}). 
To the contrary, $C$ is long range for the BTW, because
it approaches $0$ only for $t$ exceeding the
maximum number of waves in an avalanche (Fig.~\ref{fig:corrarea}), 
which we found to scale $\sim L$, 
for $L \to \infty$. For the BTW we further tested
the FSS form
\begin{equation} 
C(t,L)=t^{-\gammacorr} 
g(t/L^{D_c})\hskip 0.5 cm  \textrm{($t,L\gg 1$)},
\label{eq:FSShypothesis}
\end{equation}
on the basis of the $L \to \infty$ scaling of the 
moments~\cite{DeMenech,Tebaldi}
\begin{equation}
\ave{t^q}_L=\sum_t C(t,L) t^q\sim L^{\taucorr(q)}.
\end{equation}
FSS would imply the piecewise linear form
\begin{equation}
\taucorr(q)=\left\{
\begin{array}{ll}
D_c(q-\gammacorr+1) &\textrm{if $q\geq \gammacorr-1$}, \\
0 &\textrm{if $q< \gammacorr-1$}. \\
\end{array}
\right.
\label{eq:linearcorr}
\end{equation}
 Fig.~\ref{fig:tauqcorr} shows the extrapolated 
$\taucorr(q)$, which has an approximately linear part for 
$1\leq q \leq 4$, consistent with
$D_c=1.02\pm0.05$
and $\gammacorr=0.40\pm0.05$. The curvature for $q<1$ is due to the fact that,
for finite $L$, a logarithm ($\gammacorr=0$) can not 
be easily distinguished
from a power law with $\gammacorr \gtrsim 0$. The inset of
Fig.~\ref{fig:tauqcorr} shows this logarithmic growth 
of $\ave{t^q}_L$.
We conclude  that, for the BTW, a simple power law tail
$C(t,\infty)\sim t^{-\gammacorr}$, is
a first, rough,  approximation.
The increment $y(t)=\sum_{k=1}^t s_k$
is comparable to the trail of a fractional Brownian
motion with Hurst exponent $H=(2-\gammacorr)/2$~\cite{Sole},
such that $H=0.80\pm0.03$ 
for the BTW.
$H$ can be measured directly from the fluctuation 
\begin{equation}
F(t,L)=\left[{\ave{\Delta y(t)^2}}_L-{\ave{\Delta y(t)}_L}^2\right]^{1/2},
\end{equation}
with $\Delta y(t)=y(k+t)-y(k)$~\cite{Sole}, which should scale as 
$F(t,\infty)\sim t^H$. 
Fig.~\ref{fig:variance} reports tests of this scaling for 
$L=2048$. For the BTW, $H\sim 0.85$ at low $t$, 
in agreement
with $\gammacorr=0.40\pm0.05$.
A crossover to
$H=1/2$ is observed for large $t$. The crossover time of course 
increases with $L$. $H=1/2$ corresponds to a process
with $C$ exponentially decaying with $t$, or with
$C(t,\infty)=\delta_{t,0}$, as is the case for the
M. Thus,  the crossover is due to the fact that, beyond the
maximal time duration of avalanches, waves are 
uncorrelated. $H\neq 1/2$ implies long range correlations,
as we find for the BTW. Furthermore, $H>1/2$ corresponds
to persistency: an increasing or decreasing trend of
$y$ in the past mostly implies a similar
tendency in the future. This accounts for the observed
expansion and contraction phases in avalanche growth~\cite{Ktitarev-waves}.
The above features of $C$ should be responsible for the peculiar
long range on--site correlations of the noise expected when mapping
the BTW onto a discrete interface growth equation~\cite{Alava}. 

The BTW has been shown to display a non constant gap in
the high $q$ moments of some avalanche PDF's~\cite{Tebaldi}. Thus,
it should not surprise if, unlike assumed
in Eq.~\myref{eq:linearcorr}, also $C$
would manifest similar multifractal properties.
Indeed, a more accurate analysis shows that,
for high $q$, the gap  $d\taucorr(q)/dq$ grows slowly with $q$,
beyond the above estimate of $D_c$.
This multifractal character can be embodied in the 
more general scaling ansatz~\cite{Tebaldi}
\begin{equation}
C(t=L^\alpha,L)=L^{\fcorr(\alpha)-\alpha},\hskip 0.5 cm
\textrm{($L\gg 1$)},
\label{eq:MShypothesis}
\end{equation}
with a nonlinear singularity
spectrum $\fcorr > - \infty$ 
in an $\alpha$-interval covering the whole range of possible gaps
and linked to $\taucorr$ by Legendre transform. 
$C$ would satisfy the FFS ansatz~\myref{eq:FSShypothesis} only if 
$\fcorr$ were a linear function, i.e. 
$\fcorr(\alpha)=-(\gammacorr-1)\alpha$ if $\alpha\in [0,D_c]$,
$\fcorr(\alpha)=-\infty$ otherwise.
The FSS picture given above is in fact only an
approximation. This explains also the slight curvature of the  $F$ plot 
for low $t$, which
makes the direct measurement of $H$ ambiguous
(Fig.~\ref{fig:variance}).
Rather than attempting a more precise determination of $\taucorr$ and
$\fcorr$, below we clarify the difference between BTW and M, and the
origin of multiscaling in the former, in the light of probabilistic
concepts.

\begin{figure}[htb]
  \centerline{
  \epsfxsize=\figsize
  \epsffile{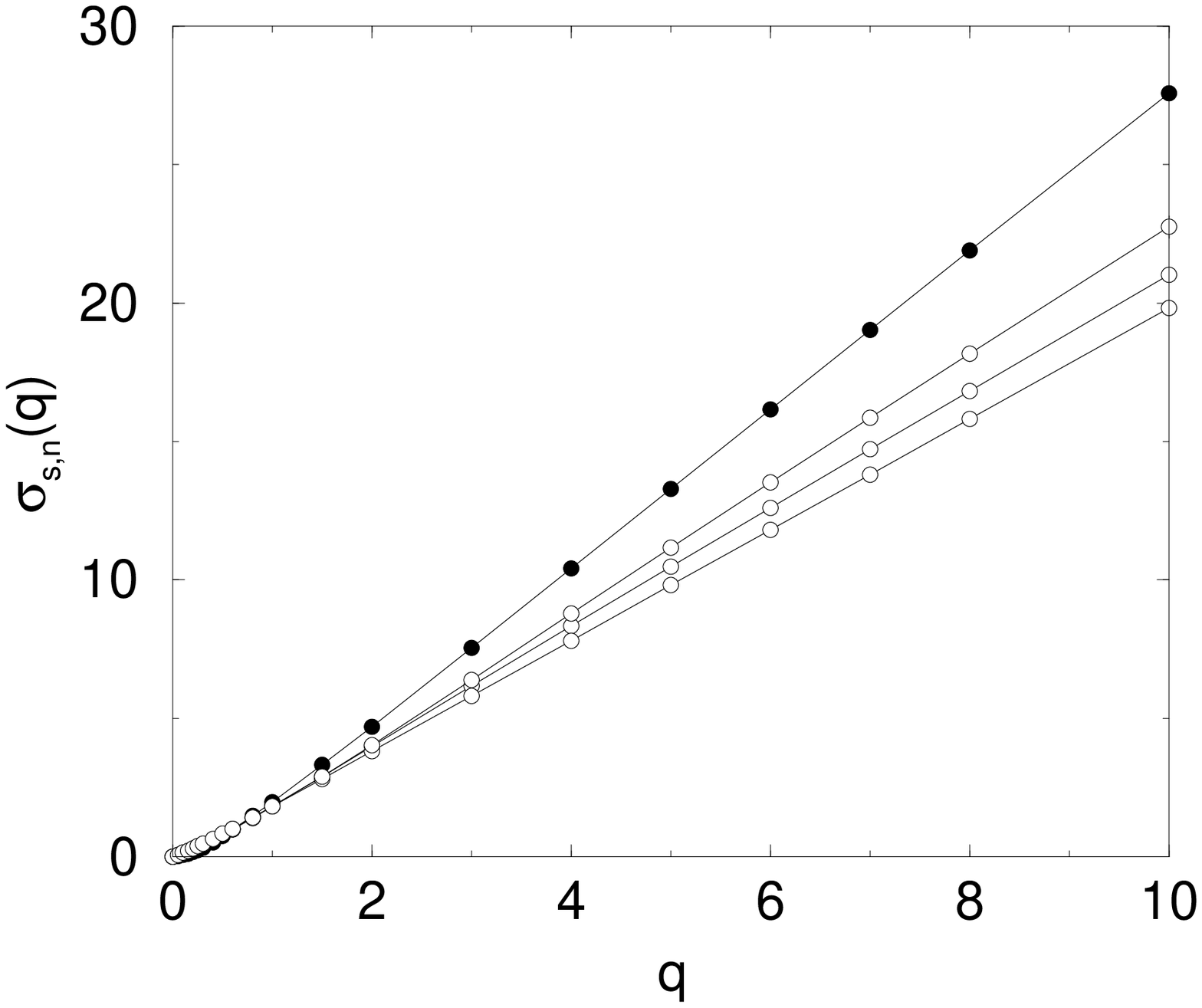}
  }
  \caption{$\tauwaves{n}(q)$ for the BTW, $n=1,8,24$ (open circles, from the
  bottom). The full circles show $\tausize(q)$.}
  \label{fig:wavessum}
\end{figure}

Waves have a relatively simple behavior. So, 
in a renormalization group (RG) spirit,
we can coarse grain the time, by looking at the PDF, $P^{(n)}(s,L)$, 
of the sum of the sizes of $n$ consecutive
waves, regardless of the avalanche they belong to.
Since avalanches are constituted by an infinite 
number~\cite{Note-nwaves} of waves for $L\to\infty$, 
by sending also $n\to\infty$, we expect $P^{(n)}$ 
to approach the PDF of avalanche sizes as a RG fixed point. 
This approach can be monitored on the basis of the effective
moment scaling exponents defined by
\begin{equation}
\ave{s^q}_{n,L}=\int ds\; s^q P^{(n)}(s,L)\sim L^{\tauwaves{n}(q)}. 
\end{equation}
For the M, $\tauwaves{n}(q)$ does not depend on
$n$, and, within our accuracy, is equal from the start to 
$\tausize(q)$, such that $\int ds\; s^q
P(s,L)\sim L^{\tausize(q)}$, with $P(s,L)$ representing the avalanche
size PDF.
To the contrary, in the BTW,
as $n$ increases, $\tauwaves{n}(q)$ varies and  
moves gradually towards the appropriate
$\tausize(q)$ (Fig.~\ref{fig:wavessum}). 

The result for the M can be explained on the basis of the fact
that PDF's of independent variables, satisfying a scaling
ansatz of the type~\myref{eq:MShypothesis}, have a spectrum
which does not change under convolution~\cite{Note-spectra}.
For example, if we put $P_w(s=L^{\alpha},L) \sim
L^{f_w(\alpha)-\alpha}$, where $f_w$ is the spectrum of the wave size PDF,
one can verify that
\begin{eqnarray}
\lefteqn{
P^{(2)}(L^{\alpha},L)=}\\\nonumber
&&\int d(L^\beta)\;
P_w(L^{\alpha}-L^{\beta},L) P_w(L^{\beta},L) \sim L^{f_w(\alpha)-\alpha}.
\end{eqnarray}
Thus, also the spectrum associated with
$P^{(n)}(s,L)$, which is the convolution of $n$ $P_w$'s, does not depend on
$n$. This implies the $n$--independence of
$\tauwaves{n}(q)$, which is determined once $\fwaves$ is given.
%$\tauwaves{n}(q)=\sup_{\alpha}\left\{
%\fwaves(\alpha)+q\alpha\right\}$. 
Now, let $P(s,m,L)$ be the probability of having an 
avalanche with $s$ topplings and $m$ waves: due to the uncorrelated
character of different wave sizes, for the M 
$P(s,m,L)$ is also the convolution of $m$ $P_w$'s, i.e 
$P(s,m,L)=P^{(m)}(s,L)$. 
Therefore, if $P(m,L)$ is the PDF of the total number, $m$, of waves in an 
avalanche, one has
\begin{equation}
P(s,L)=
\sum_{m} P^{(m)}(s,L)P(m,L),
\label{eq:sumPDF}
\end{equation}
such that
\begin{equation}
L^{\tausize(q)} \sim \sum_m L^{\tauwaves{m}(q)} P(m,L),
\label{eq:tauS-tauSn}
\end{equation}
which, together with the above results, imply 
$\tauwaves{n}(q)=\tausize(q)$. The
coarse graining of waves in the M
does not modify the block PDF spectrum,
which is from the start at its fixed point, representing
also the scaling of avalanches.

In the BTW case,
the nontrivial composition of correlated waves is responsible for the
change with $n$ of the effective singularity spectrum of $P^{(n)}$
(Fig.~\ref{fig:wavessum}).
For $n=1$,  the simple linear FSS 
form  $f_w(\alpha)=-(\tau_w-1)\alpha$ ($\alpha\in[0,D_w]$) applies, 
while for $n\to\infty$ one should recover the nonlinear form 
needed to describe the
complex scaling of $P(s,L)$~\cite{Tebaldi}. 
In practice, sampling limitations
prevent us from reaching very high $n$. However,
even if convergence is relatively slow,
the tendency of $\tauwaves{n}(q)$ to move towards
$\tausize(q)$ is very manifest. 
We could clearly detect the increase with $n$ of the  
gap $d\tauwaves{n}(q)/dq$,  at fixed high $q$: 
for example, we determined
$d\tauwaves{n}(4)/dq\simeq2.00,2.20,2.25,2.38$, for $n=1,8,12,24$,
respectively. Furthermore, while  the asymtoptic gap for
$P_w(s,L)$ gets readily to the maximum value, i.e.
$d\tauwaves{1}(q)/dq=2$ for  $q\geq 1$, as soon as $n>1$  a
constant gap could not be detected for $\tauwaves{n}(q)$; this confirms
the progressive appearance of effective multifractal scaling  
for the block variable.  

In summary, a comparative
analysis of wave time series shows
that the different forms of universal scaling
in the BTW and M are determined by distinct dynamical
correlation patterns. For the M case,  the
wave level of description is in fact coarse grained enough to
fully account, without further modifications, also for the avalanche level.
Waves are uncorrelated and their PDF, satisfying FSS, 
coincides with the avalanche PDF, as far as exponents
are concerned. To the contrary, in the BTW, under coarse graining, long time 
correlations substantially modify the scaling properties of waves, 
determining also multiscaling
features. We regard these as prototype
mechanisms to be generally expected in sandpile
and similar SOC models. The 2D BTW behavior is probably less
generic than the M one. For example, in the 3D
BTW, the fact that avalanches are constituted by
an average number of waves which remains finite for $L\to\infty$, 
 strongly suggests
a M type  mechanism and FSS for the avalanche PDF.
Indeed, even if subsequent waves are not strictly
uncorrelated, like in the 2D Manna model,
$C$ should decay extremely fast, with an $L$ independent time cutoff. 
Thus, coarse graining in time
can not substantially modify the
PDF of block variables, with respect to $P_w$.
We expect the exactly known wave exponents to
apply also to the avalanche size PDF. 
Indeed, numerically determined avalanche exponents for 3D BTW turn out to be
strikingly close to the wave values~\cite{Ktitarev-Lubeck}.

We thank C. Tebaldi and C. Vanderzande for useful discussions.
Partial support from the European Network Contract 
No. ERBFMRXCT980183 is acknowledged. 

%%%%%%%%%%%%%%%%%%%%%%%%%%%%%%%%%%%%%%%%%%%%%

%%%%%%%%%%%%%%%%%%%%%%%%%%%%%%%%%%%%%% FIGURES 


\end{document}